# A NEW FIRST-PRINCIPLES CALCULATION OF FIELD-DEPENDENT RF SURFACE IMPEDANCE OF BCS SUPERCONDUCTOR AND APPLICATION TO SRF CAVITIES

B. P. Xiao[1#] and C. E. Reece[2]
[1] Brookhaven National Laboratory, Upton, New York 11973
[2] Thomas Jefferson National Accelerator Facility, Newport News, Virginia 23606

*Abstract*

There is a need to better understand the intrinsic limit of radiofrequency (RF) surface impedance that determines the performance of superconducting RF cavities in particle accelerators. Here we present a field-dependent derivation of Mattis-Bardeen (M-B) theory of the RF surface impedance of BCS superconductors based on the shifted Density of States (DoS) resulting from coherently moving Cooper pairs [1]. The surprising reduction in resistance with increasing field is explained to be an intrinsic effect. Using this analysis coded in Mathematica™, survey calculations have been completed which examine the sensitivities of this surface impedance to variation of the BCS material parameters and temperature. Our theoretical prediction of the effective BCS RF surface resistance ($R_s$) of niobium as a function of peak surface magnetic field amplitude agrees well with recently reported record low loss resonant cavity measurements from Jefferson Lab (JLab) and Fermi National Accelerator Lab (FNAL) with carefully, yet differently, prepared niobium material. The results present a refined description of the "best theoretical" performance available to potential applications with corresponding materials.

## INTRODUCTION

Superconducting radiofrequency (SRF) accelerating cavities for particle accelerators made from bulk niobium (Nb) materials are the state-of-art facilities for exploring frontier physics. The quality of the SRF cavities is characterized by the so-called quality factor $Q$ under different peak magnetic field on the cavity inner surface $B_{pk}$, with $Q=G/R_s$ and $G$ the geometry factor of the cavity, which is cavity design dependent.

Remarkable results have been achieved in SRF cavity performance: for a single-cell re-entrant shape cavity at Cornell University, the maximum accelerating gradient has been pushed to 197.1 mT $B_{pk}$ with quality factor ($Q_0$) higher than $10^{10}$ at 1.3 GHz and 2.0 K temperature [2], shown as red square ■ in Figure 1; and for a single-cell TESLA shape fine grain (FG) cavity TE1AES011 with surface doping with nitrogen at 800°C (HT-N) by FNAL, the cavity exhibits a $Q_0$ approaching $1\times10^{11}$ with 80 mT magnetic field at 1.3 GHz and 2.0 K, limited by quench at 127 mT [3], shown as green dot ● in Figure 1. In Error! Reference source not found. we also show the test result for a 7-cell Continuous Electron Beam Accelerator Facility (CEBAF) upgrade prototype cavity LL002 with >300 μm buffer chemical polishing (BCP) surface treatment surface treatment in black triangle ▲, which is recently considered to be a typical $Q$ curve, including low field $Q$ increase, middle field $Q$ decrease and one type of high field $Q$ drop[4].

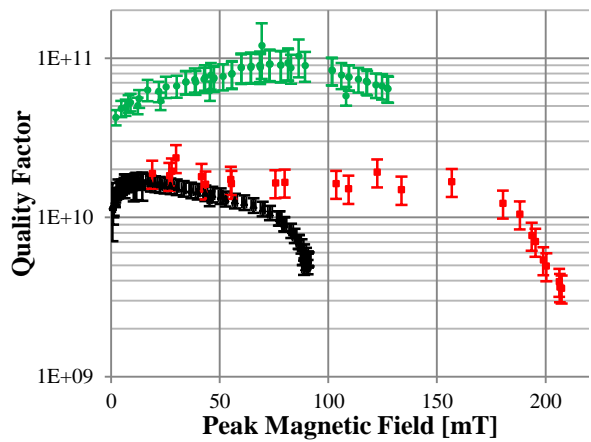

Figure 1: Cavity performance at 2 K for: ▲ 1.5 GHz 7-cell LL002 cavity, ■ 1.3 GHz Cornell single-cell re-entrant shape cavity, and ● 1.3 GHz single cell TESLA cavity with HT-N doping. Errors on fields are small and are not shown here.

Theories are needed to explain the measured curves shown in Figure 1: the limitations on the magnetic field, the highest $Q$ that can be achieved in niobium SRF cavities [5], and the $Q$ changes with $B_{pk}$, etc. One would also like to extend the theoretical understanding to the corresponding limitations on alternative materials for possible SRF applications.

To predict the highest achievable SRF $B_{pk}$, a theory was developed to qualitatively calculate the upper limit of the magnetic field in which the Meissner state can exist as a metastable state based on the energy barrier at the surface that impedes the penetration of vortices into the bulk, the so-called superheating field theory [6].

To explain the $Q$ in the low field limit, for example, the $Q$ value at $B_{pk}=0$ in Figure 1, M-B theory was developed to calculate the surface impedance of conventional superconductors at high frequency and low temperature [7].

The RF surface impedance of a superconductor may be considered a consequence of the inertia of the Cooper pairs. The resulting incomplete shielding of RF field allows the superconductor to store RF energy inside its surface, which may be described by a surface reactance, $X_s$. The RF field that enters the superconductor interacts with quasi-particles, causing power dissipation, described by a surface resistance, $R_s$. M-B theory started from the BCS theory [8], using the quasi-particle states (electron above Fermi level and hole below Fermi level) distribution at 0 K and probability of occupation at $T < T_c$. The single-particle scattering operator was calculated and applied into anomalous skin effect theory to obtain the surface impedance. M-B theory, however, does not consider the field dependence of surface impedance. In particular, its real part, surface resistance, which is of great interest in SRF applications, is unaddressed.

To attempt to explain the $Q$ changes with $B_{pk}$, several theories have been developed trying to address aspects of the experimentally observed behavior of SRF cavity performance. A summary of these theories was assembled by Visentin [9]. These theories, however, do not consider changes to the low field limit assumption in BCS and M-B theories, and do not address a theoretical limit for the quality factor as a function of the magnetic field amplitude.

Recently a new model has been put forward by Xiao *et al.*, [1] starting from the BCS theory with a net current in a superconductor by taking a pairing ($k_1\uparrow$, $k_2\downarrow$), $k_1$ and $k_2$ the wave vectors of the particles, ↑ spin up and ↓ spin down, with $k_1+k_2 = 2q$, and $2q$ the same for all virtual pairs [8], the particle states distribution at 0 K were calculated, together with the probabilities of particle occupation with finite temperature and subsequently applied to anomalous skin effect theory, to obtain a new derivation of RF field dependence of the surface impedance of a superconductor.

A Mathematica™ program has been developed by Xiao to accomplish the calculation of the resulting challenging quadruple integral. It is applicable to any standard superconductor described by BCS theory. The code reproduces the heretofore standard M-B theory result at zero field as calculated, for example, by the commonly-used Halbritter code, SRIMP. [10]

A rather surprising result of the calculation, with significant importance to SRF applications, is the prediction of non-linear, *decreasing* surface resistance in an RF field regime that is prime domain for accelerator applications. The corresponding prediction of *increasing* $Q_0$ with field matches remarkably well recent reports of record-breaking low losses [11, 12] and raises the prospect that the common expectation of "best theoretical" cryogenic performance from Nb, and in principle other BCS superconductors, may be dramatically revised for the better.

We have used this code to perform a parametric sensitivity survey with each the characteristic BCS material parameters of the field-dependent RF surface impedance in hopes of supporting increased insight into a performance optimization strategy.

## FIELD DEPENDENT EXTENSION OF BCS AND M-B THEORY

In the BCS theory, paired particles in the ground state, with total mass $2m$ and zero total momentum that occupy state ($k\uparrow$, $-k\downarrow$), with velocity $V_k$ in random direction, and energy relative to the Fermi level $\varepsilon_F$ of $|\varepsilon_k|$, have been considered to give minimum free energy for superconductors. For notation, we refer to Fermi velocity as $V_F$ and Fermi momentum as $P_F$.

In the extended theory, states with net flow in a certain direction can be obtained by taking a pairing ($k+q\uparrow$, $-k+q\downarrow$), with total momentum $2q$ the same for all Cooper pairs, corresponding to net velocity $V_s = \hbar q/m$. This change may be illustrated by a slice of the Fermi sphere depicted in Figure 2.

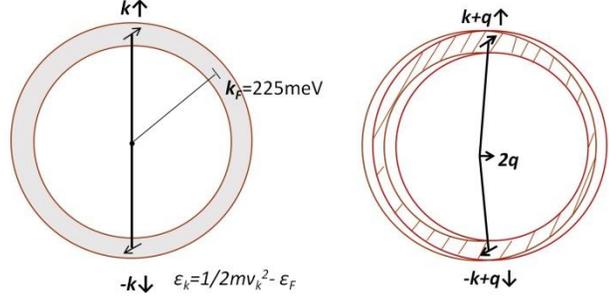

Figure 2. Slice of Fermi sphere of the superconductor: in the low field limit BCS theory (left), and with net momentum $2q$ the same for all Cooper pairs in the extended theory (right). Numbers labeled are typical Nb parameters [13].

*Change in energy*

In the original BCS theory, the Bloch energy $|\varepsilon_k|$ (relative to $\varepsilon_F$) of the particle, corresponding to the Bloch energy $\varepsilon_k$ (relative to $\varepsilon_F$) of the electron before condensation, will change to $E_k = \sqrt{\varepsilon_k^2 + \Delta^2}$ after condensation, with $\Delta$ the energy gap, as shown in Figure 3(a). Two particles (fermions) in the same energy state $k$ nearby the Fermi level $\varepsilon_F$, with one ↑ and the other one ↓, can be attracted to each other via electron-phonon interaction and become a Cooper pair (boson), with the energy of the boson reduced to zero, shown as the black line on the bottom of Figure 3(a). The minimum energy needed to break a Cooper pair is $2\Delta$.

With the theory extension, the Bloch energies for two particles that are going to combine into one Cooper pair ($k+q\uparrow$, $-k+q\downarrow$) after condensation are no longer the same; they split into two different Bloch energies, $\varepsilon_{k+q\uparrow} = \varepsilon_k + \varepsilon_s + \varepsilon_{ext}$ for ↑ and $\varepsilon_{-k+q} = \varepsilon_k + \varepsilon_s - \varepsilon_{ext}$ for ↓ before condensation, where $\varepsilon_s = mV_s^2/2$, $\varepsilon_{ext} = p_F V_s x$ and $x = \cos\alpha$, and $\alpha$ is the angle between $V_s$ and $V_F$. Even though the absolute value of $V_s$ is much smaller than that of $V_F$, the angle $\alpha$ between these two velocities significantly affects the Bloch energies for the particles. The energies after condensation change to $E_{k+q\uparrow} = E_k + \varepsilon_{ext}$ for spin up and $E_{-k+q\downarrow} = E_k - \varepsilon_{ext}$ for spin down, with $E_k = \sqrt{(\varepsilon_k + \varepsilon_s)^2 + \Delta^2}$, shown as equations (4) and (5) in [1]. In Figure 3(b) the particle energies after condensation $E_{k+q\uparrow}$ and $E_{-k+q\downarrow}$ as a function of $\varepsilon_k$ are shown for the specific case of $\varepsilon_{ext} = 0.4\Delta$. The minimum energy needed to break a Cooper pair remains $2\Delta$, with $\Delta + \varepsilon_{ext}$ for ↑ and $\Delta - \varepsilon_{ext}$ for ↓, also shown in Figure 3(b). Detailed angle-averaged calculation also shows a slight decrease in effective $\Delta$ with increasing $V_s$ as illustrated below.

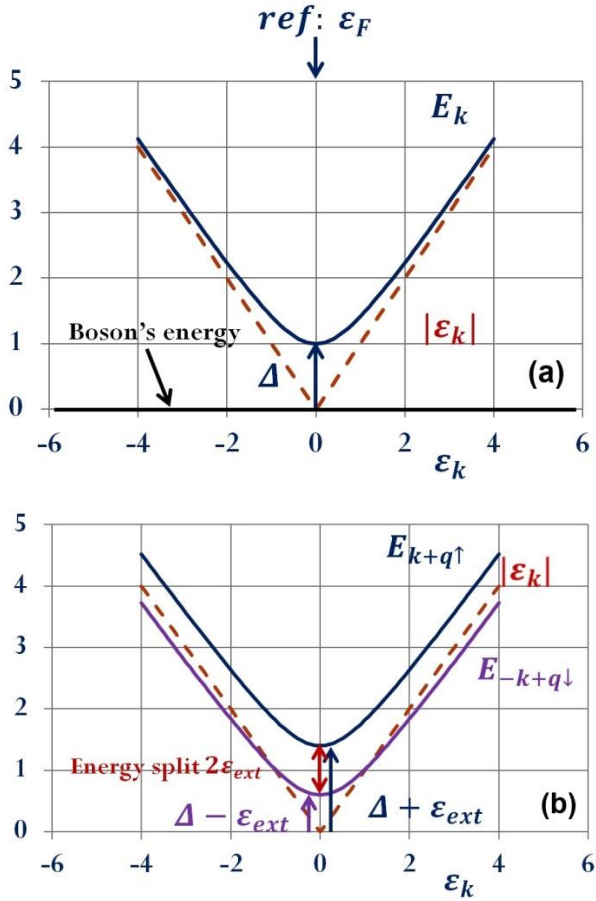

Figure 3. Top (a): Particle energy before condensation, $|\varepsilon_k|$, and after condensation, $E_k$, as a function of electron energy $\varepsilon_k$ relative to Fermi energy $\varepsilon_F$. The energy of Cooper pairs is zero, and the minimum energy needed to break a Cooper pair is $2\Delta$. Bottom (b): Particle energies after condensation $E_{k+q\uparrow}$ and $E_{-k+q\downarrow}$ as a function of $\varepsilon_k$ with $\varepsilon_{ext} = 0.4\Delta$. $|\varepsilon_k|$ is shown for reference only. The minimum energy needed to break a Cooper pair is $2\Delta$, with $\Delta + \varepsilon_{ext}$ for $\uparrow$ and $\Delta - \varepsilon_{ext}$ for $\downarrow$. All numbers are normalized to $\Delta$. The effect of $\varepsilon_s$ (small compared with $\Delta$) is not considered here.

*Change in DoS and distribution function*

The DoS $N(E)/N_0$ and the distribution function $f$ as a function of the Bloch energy derived from BCS theory are shown in Figure 4(a) with $T/T_c$ of 0.97, similar to Figure 1 in [14]. Please note $E$ equals to 0 at the dashed line, with its number to be positive on both sides, referring to holes on the left and electrons on the right. In the extended theory, the modified DoS and the probability of occupation at $T<T_c$, with their angle integrations shown as equations (21) and (20) in [1], respectively, are both angle dependent with $\alpha$. The angle dependence of the modified DoS and the probability of occupation at $T<T_c$ as a function of the Bloch energy is depicted in Figure 4(b), and also in Figure 4(c) with angle integration. In these plots Cooper pair net momentum is chosen to be $V_s = 0.4\Delta/P_F$ for illustration. Since holes are counted on the left and electrons are counted on the right, there is a sharp change at $E = 0$ for the distribution function $f$.

From the Figures one can observe that even though in the average, the gap is reduced by a value of $P_F V_s$, the energy that is needed to separate the particles in a Cooper pair does not change significantly with α changes, remaining $2\Delta$ as illustrated in Figure **3**(b). If the tunnelling effect were used to measure the gap in this flowing-current situation, which actually measures the gap in the quasi-particle distribution, the result would show a value of $2(\Delta - P_F V_s)$; whereas, if infrared photons were used to measure the energy required to break the Cooper pairs, the value would be $2\Delta$.

*Change in M-B theory*

In the original M-B theory [7], the single particle scattering matrix was calculated using the modified DoS and the probability of occupation at $T<T_c$, and then applied to the anomalous skin effect theory to derive the surface impedance of superconductors.

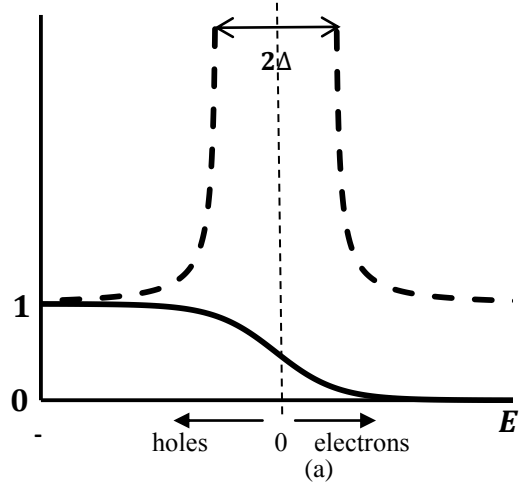

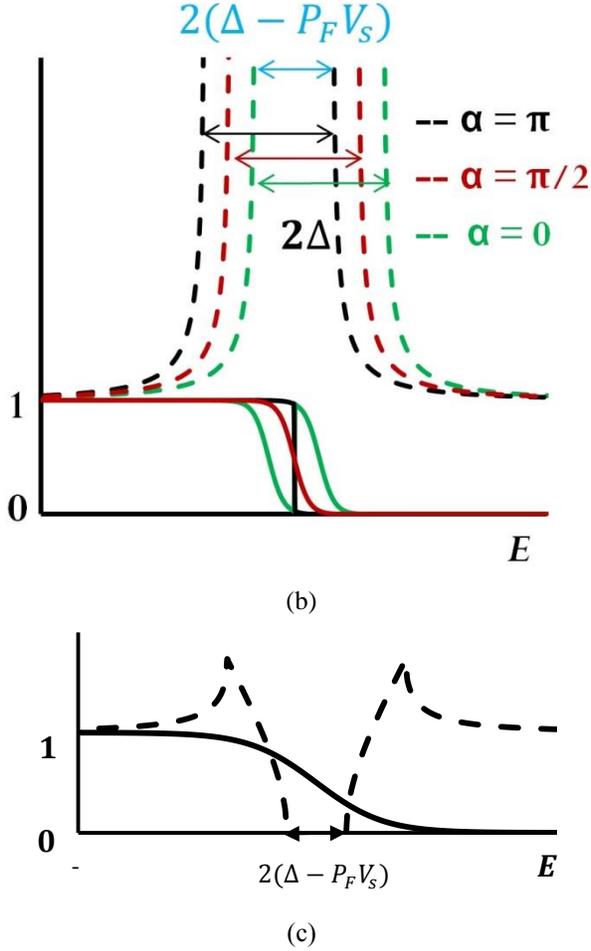

Figure 4. (a) DoS (dotted curve) and distribution function (solid curve) in the low field limit; (b) DoS (dotted curve) and distribution function (solid curve) with moving Cooper pairs, angle-dependent. (c) DoS (dotted curve) and distribution function (solid curve) with moving Cooper pairs, angle averaged; Plotted with $T/T_c=0.97$, and with $P_F V_s = 0.4\Delta$ for (b) and (c).

In the extended theory, the changes in the modified DoS and the probability of occupation cause a significant change in the single particle scattering operator [1, 8], which leads to a field dependence of $R_s$. The detailed calculation has been shown in [1].

While the numerical calculation of the surface impedance in the M-B theory is complex [14], equation (3.5) in [7] could be relatively simple in the extreme anomalous limit, as shown in (3.9) and (3.10) of [7]. For SRF applications in the low field limit, the surface resistance $R_s$ simplifies to [14]:

$$R_s \propto 2 \int_\Delta^\infty [f(E) - f(E + \hbar\omega)] g(E) dE \quad (1)$$

with $\hbar\omega$ being the photon energy and $g(E) = \frac{E_1 E_2 + \Delta^2}{\varepsilon_1 \varepsilon_2}$, a function related to the modified DoS, with $E_1 = E$ and $E_2 = E + \hbar\omega$.

The expression of $R_s$ is similar to equation (12) in [14] deduced from the Golden Rule. The dynamic balance in the photon absorption and emission between $E_1$ and $E_2$ causes net power dissipation.

In the extended theory, the calculation of the surface impedance is even more complex than in the M-B theory. In the extreme anomalous limit, the surface resistance in this extended theory changes to:

$$R_s \propto 2 \int_{\max(\Delta-\varepsilon_{\text{ext}},\Delta-\varepsilon'_{\text{ext}}-\hbar\omega)}^\infty [f(E_1) - f(E_2)][f(\varepsilon_{\text{ext}}) + f(-\varepsilon_{\text{ext}})] g(E,\alpha,\alpha') dE. \quad (2)$$

with $E_x = E_2$ for $\varepsilon_{ext} < \varepsilon'_{ext} + \hbar\omega$ and $E_x = E_1$ for $\varepsilon_{ext} > \varepsilon'_{ext} + \hbar\omega$, with $E_1 = E + \varepsilon_{ext}$, $E_2 = E + \varepsilon'_{ext} + \hbar\omega$, and $\varepsilon_{ext} = P_F V_s \cos\alpha$ being the additional energy from the energy split, and $\varepsilon'_{ext}$ that for another particle state with different angle $\alpha'$, and $g(E,\alpha,\alpha') = \frac{E_1 E_2 + \Delta^2}{(\varepsilon_1+\varepsilon_s)(\varepsilon_2+\varepsilon_s)}$, a function related to the modified DoS.

## CALCULATION RESULTS AND INTERPRETATION

While the extended theory is general to any BCS material, the present analysis focuses on niobium, using the following characteristic parameters as standard conditions: $\Delta_0/kT_c(0)=1.85$, $T_c(0) = 9.25$ K, $\xi_0 = 40$ nm, $\lambda_L(0) = 32$ nm, and mean free path $\iota = 50$ nm [13], exploring the predicted surface impedance variation with departures from these values. The calculated standard condition surface impedance of niobium at 1.3 GHz and 2.0 K is shown as a function of Cooper pair velocity in Figure 5, one may refer to [1] for similar results at 1.5 GHz.

Beginning with a 1.3 GHz 0 m/s Cooper pair velocity $V_s$ static case surface resistance at 2.0 K of 8.4 n$\Omega$, $R_s$ first decreases with increasing $V_s$, then increases, with a minimum $R_s$ of 1.5 n$\Omega$ at 200 m/s. Since the supercurrent density varies both with depth into the surface and time within the RF cycle, the surface resistance does as well, so calculation of an effective surface resistance must integrate over both material depth and RF cycle. See reference [1] for discussion of simplifying assumptions that are made in the analysis.

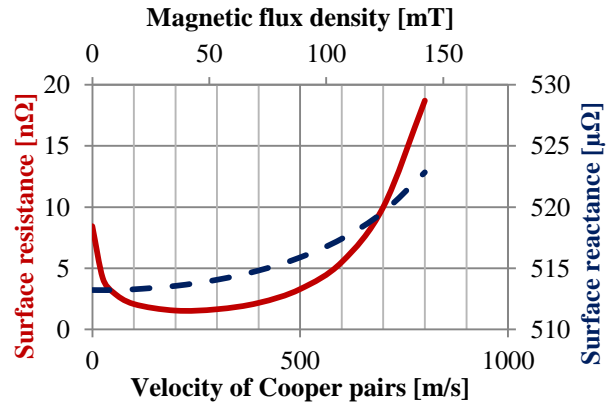

Figure 5. Surface resistance (red line) and reactance (blue dashed line) versus Cooper pair velocity for Nb at 2 K at 1.3 GHz.

The resulting predicted effective surface resistance under the standard Nb parameter conditions is shown in Figure 6, together with the result of similar calculations at 0.7 and 0.4 GHz.

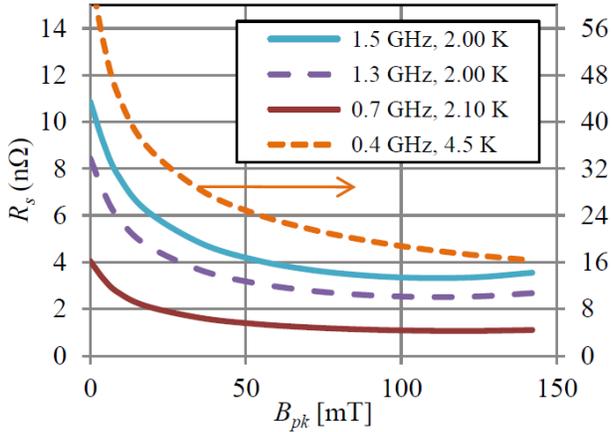

Figure 6. Calculated effective surface resistance under "standard conditions" for Nb versus peak RF magnetic field for 1.5 and 1.3 GHz at 2.0 K, 0.7 GHz at 2.1 K, and 0.4 GHz at 4.5 K.

Due to the field dependence of $R_s$, the field distribution inside a cavity can yield a non-uniform $R_s$, even with uniform temperature distribution on the cavity's inner surface. Four different cavity shapes have been evaluated: TeV-Energy Superconducting Linear Accelerator (TESLA) shape 9-cell cavity [15], TESLA shape single cell cavity (the end cell of the TESLA shape 9-cell cavity), CEBAF high gradient (HG) 7-cell cavity [16] and CEBAF C100 LL cavity [17]. Less than 0.1 nΩ deviation from the "standard condition" data shown in Figure 6 was found for all shapes. This would not be the case for the more complex structures typically used for low-$\beta$ accelerator applications.

To understand the change of $R_s$ under different $B$, one may start from a single particle scattering analysis. From the description in references [7, 8], with at least one single particle in either initial state or final state, one particle has different possibilities to transition from one energy state $E$ with any arbitrary number, to another energy $E'$, associated with either absorbing or releasing one photon. The net effect here is absorbing photons and releasing thermal energy, illustrated in the top of Figure 7. One should note that the scattering procedure should be considered as a quantum procedure, and energy conservation should be considered in the overall effect.

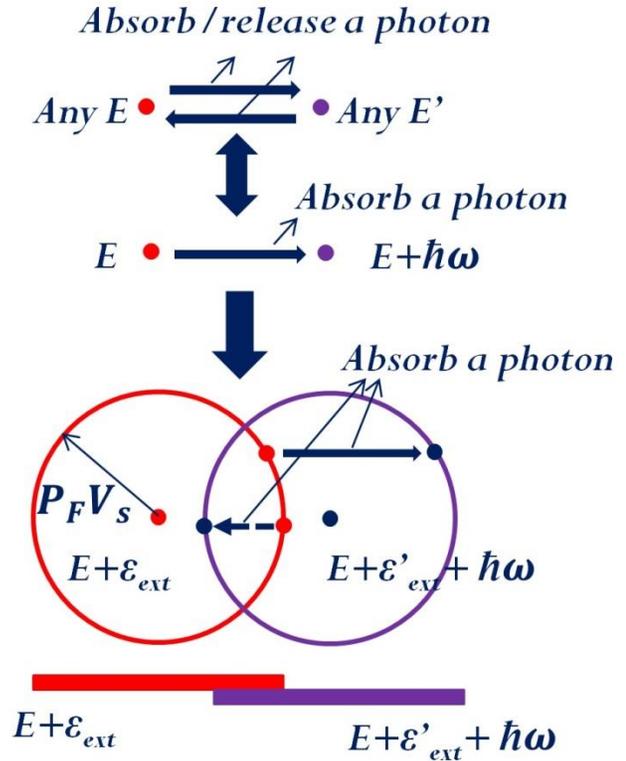

Figure 7: Energy consumption procedure of quasi-particles: described in the original BCS and M-B theory in the low field limit (top), mathematically equivalent description at low field limit (middle) and mathematically equivalent description with net momentum $2q$ the same for all Cooper pairs (bottom).

The net effect of the above procedure is mathematically equivalent to the following: one particle, with any arbitrary energy $E$, that could jump to a higher energy $E + \hbar\omega$ with absorption of one photon, also has a certain possibility to jump back from $E + \hbar\omega$ to $E$ and release one photon, with the net effect to be some probability of jumping from $E$ to $E + \hbar\omega$ and absorbing a photon, as shown in the middle of Figure 7.

In the extended theory with an angle between $V_F$ (which will be in any random direction) and $V_S$, the Bloch energy for two particles in a Cooper pair splits, and an angle dependence appears, the energy consumption which corresponds to the transition between two fixed modified energy states $E_1 = E$ and $E_2 = E + \hbar\omega$ in the low field limit changes to between $E_1 = E + \varepsilon_{ext}$ and $E_2 = E + \varepsilon'_{ext} + \hbar\omega$. The energy spread caused by the angle between $V_F$ and $V_S$, shown as the red and purple circles in the bottom of Figure 7 as a function of angle, projected to energy space appears as red and purple bars, affects the energy levels in the distribution function, as well as the DoS embedded in the Golden Rule. The net effect equals to some probability of a particle scattering from a point on the red circle/bar, to any point on the purple circle/bar satisfying energy conservation, associating with absorbing one photon. Same as the previous analysis, the scattering procedure should be considered as a quantum procedure, and energy conservation should be considered in the overall effect.

A consequence appears to be attractive: In the extended theory, the net effect that equals to the scattering associating with photon absorption from $E + \varepsilon_{ext}$, the red circle/bar in the bottom of Figure 7, to $E + \varepsilon'_{ext} + \hbar\omega$, the purple circle/bar in the bottom of Figure 7, is not always from a lower energy state to a higher energy state with energy difference to be a photon energy. A quasiparticle may scatter from a lower energy state to a higher energy state with energy difference less than a photon energy, or even a higher energy state to lower, together with the absorption of a photon. shown as the dashed arrow in the overlapped region of the red and purple energy bars. This overlapped region could be significant since $P_F V_s >> \hbar\omega$ could occur for SRF applications with typical photon energy. This process "borrows" energy from those scatterings from low energy to high energy with energy difference more than a photon energy and causes cancellation effect on power consumption, and mathematically, the overall effect gives a reduction in power dissipation compared to the low field limit case. The net effect gives mathematically reduced power dissipation, thus a positive yet decreasing $R_s$ appears with field increasing up to a certain level.

In order to understand the reduction of the surface resistance with increasing field up to a certain level, it is necessary to compare expressions (1) and (2), and analytically show a reduction of $R_s$ with increasing $V_s$.

It is hard to directly compare these two expressions since the lower limit of the integration is different. Now we consider that at field level just above zero, $V_s$ is a small number such that $|2p_F v_s| < \hbar\omega$. In this case $\Delta - \varepsilon'_{ext} - \hbar\omega < \Delta - \varepsilon_{ext}$ and

$$R_s \propto 2\int_{\Delta-\varepsilon_{ext}}^{\infty} [f(E_1) - f(E_2)][f(\varepsilon_{ext}) + f(-\varepsilon_{ext})]g(E,\alpha,\alpha')dE$$

At this point we change the integration from $E$ to $E_1$, so the above expression changes to:

$$R_s \propto 2\int_{\Delta}^{\infty} [f(E_1) - f(E_1 + \varepsilon'_{ext} - \varepsilon_{ext} + \hbar\omega)][f(\varepsilon_{ext}) + f(-\varepsilon_{ext})]g(E_1 - \varepsilon_{ext}, \alpha, \alpha')dE_1 \quad (3)$$

Expression (3) and expression (1) now have the same range of integration and can be directly compared. Now we evaluate the change brought by the single particle distribution function:

$$1/4 \int_{-1}^{1}\int_{-1}^{1} [f(E_1) - f(E_1 + \varepsilon'_{ext} - \varepsilon_{ext} + \hbar\omega)]dx\,dx'$$
$$= f(E_1) - [\frac{sinh(p_F v_s)}{p_F v_s}]^2 f(E_1 + \hbar\omega)$$

with $x = \cos\alpha$ and $x' = \cos\alpha'$.

The expression $sinh(p_F v_s)/p_F v_s$ is increasing with increasing $V_s$, thus with increasing $V_s$, the $R_s$ reduces, and the reduction comes from the angle-dependent modified single particle distribution function providing on average reduced opportunities for transitions.

One should note the above analysis is true only at low field. At higher fields where $|2p_F v_s| > \hbar\omega$ may occur, similar conclusions can be drawn via numerical analysis.

## PARAMETER SURVEY

In order to potentially use the observed field dependence of the surface resistance to gain insight into changes of the superconducting material parameters, we have undertaken a calculation parametric survey to assess the sensitivity of the derived RF $R_s$ to variation from our "standard parameter" set. For all conditions considered, $T_c$ is treated as fixed at 9.25 K.

Calculated values of effective $R_s$ for Nb at 1.3 and 1.5 GHz as a function of peak RF magnetic field for several temperatures between 1.5 and 2.3 K are presented in Figure 8 and 9.

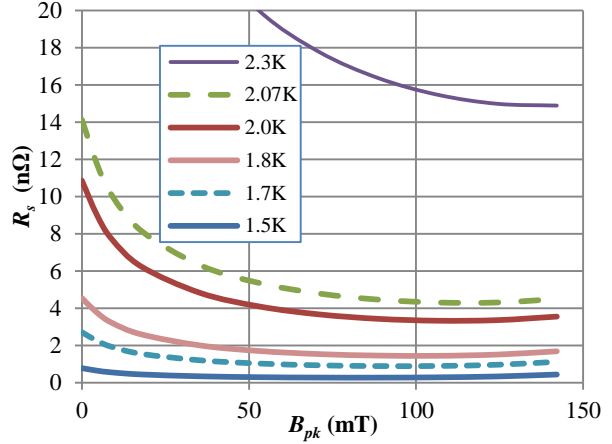

Figure 8. Effective 1.5 GHz surface resistance of standard Nb material parameters at various temperatures of interest.

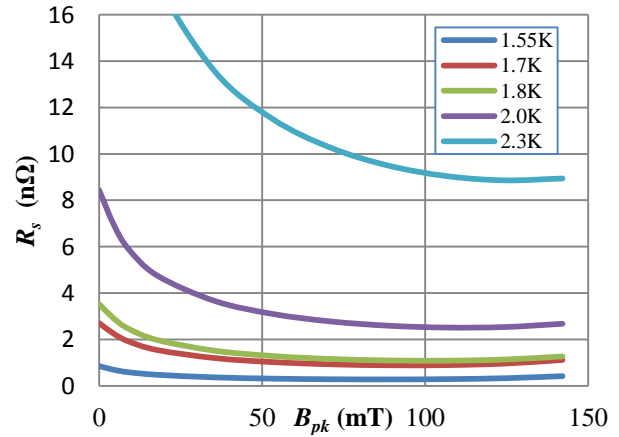

Figure 9. Effective 1.3 GHz surface resistance of standard Nb material parameters at various temperatures of interest.

The derived field dependence of $R_s$ at 2.0K with variations around coherence length and London penetration depth values of $\xi_0 = 40$ nm and $\lambda_L(0) = 32$ nm were calculated, and their deviations from the $R_s$ with "standard parameter" set are presented in Figure

10**Error! Reference source not found.** and Figure 11, respectively. Note that $R_s$ is predicted to decrease slightly more quickly with higher $\xi_0$, but is rather insensitive to $\xi_0$ in the $B_{pk}$ = 100–120 mT range, while monotonically decreasing with lower $\lambda_L$.

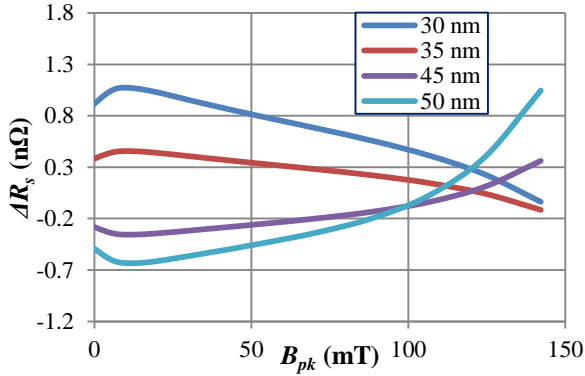

Figure 10. Deviation of effective 1.5 GHz surface resistance of Nb at 2.0 K with variations of coherence length $\xi_0$, using the "standard parameter" $\xi_0$ = 40 nm as the baseline.

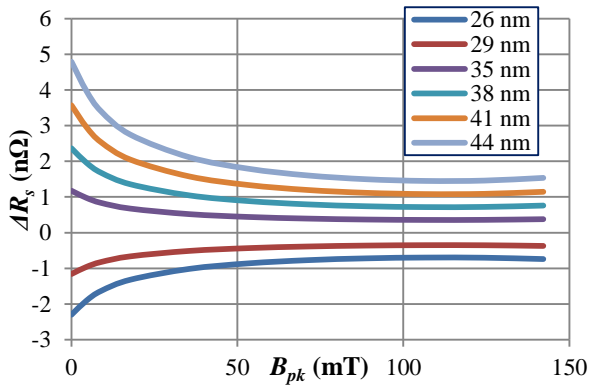

Figure 11. Deviation of effective 1.5 GHz surface resistance of Nb at 2.0 K with variations of London penetration depth $\lambda_L$, using the "standard parameter" $\lambda_L$=32 nm as the baseline.

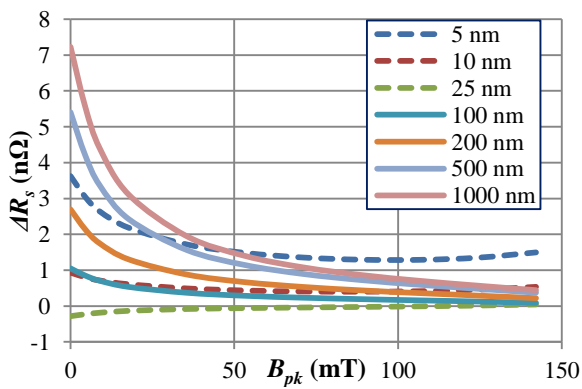

Figure 12. Deviation of effective 1.5 GHz surface resistance of Nb at 2.0 K with variations of electron mean free path $\iota$, using the "standard parameter" $\iota$ = 50 nm as the baseline.

Sensitivity of $R_s$ field dependence with electron mean free path, $\iota$, is more complex, with a clear minimum of both absolute and field-dependent components observed between 25 and 50 nm, as may be observed from Figure 12. This is consistent with data reported from previous experimental studies.[18]

Variation of the BCS gap energy yields a predicted decrease in $R_s$ with increasing gap, as expected, but fractional $R_s$ change with field shows no additional structure, as shown in Figure .

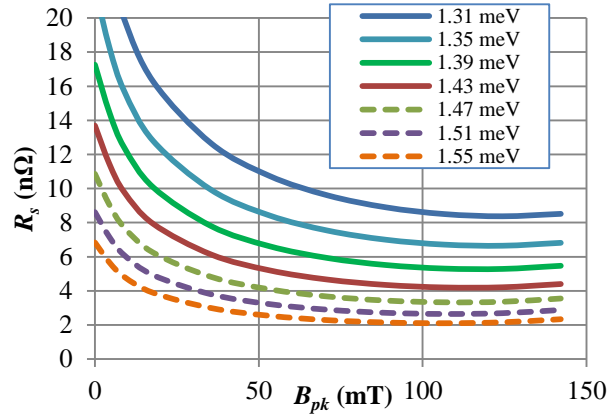

Figure 13. Effective 1.5 GHz surface resistance of Nb at 2.0 K with variations of the BCS energy gap, $\Delta_0$.

## COMPARISION BETWEEN CALCULATION AND EXPERIMENTAL DATA

Recent investigations into Nb material treatment processes which yield higher $Q_0$ of SRF accelerating cavities have begun to produce results which show increasing $Q$ with field well beyond the range of the familiar, but enigmatic "low-field $Q$ slope." [11, 12] Seeking to evaluate the relevance of the present theory to this experimental phenomenon, we plot in Figure **14**Figure  the standard calculation from Figure 5 together with the published data for four cavities, a single-cell large grain (LG) original CEBAF cell shape cavity G1G2 (LG) with 3 h 1400°C baking in JLab [19]; three single-cell TESLA shape FG cavities TE1AES003, 005, and 011, after subtracting a field-independent 1.7 nOhm from the experimental data forthe first three, and none from the fourth. The conversion to $R_s$ assumes that $B_{pk}/E_{acc}$ = 4.31 for the TE1AES003, -005, and -011 FG cavities.

Figure 14. Field-dependent BCS surface resistance at 2.0 K, calculated by Xiao's code and recent very low loss cavity test data from JLab at 1.5 GHz and FNAL at 1.3 GHz prepared by different methods. For the experimental data, ~20% error on $R_s$ and ~5% error on $B_{pk}$ are not shown here.

The calculations and the experimental results for four representative cavities shown above, exhibit a corresponding increase in $Q$ with field well beyond the range of the familiar "low-field $Q$ slope" at <20 mT, to a value of ~80 mT.

A common way to deal with the temperature and field dependence of the experimental $R_s$ is to fit the $R_s$ under the same $B$ condition by using $R_s(T)=Aexp(-U/kT)+R_{res}$. The parameters $A$, $U$ and $R_{res}$ thus become functions of $B$ [12, 20]. To compare the experimental fitting results of $A(B)$ and $U(B)$ shown in [20] with Xiao's extension, expression $R_s(T)=Aexp(-U/kT)$ was used to fit the calculation results shown in Figure 9, with fitting results shown in Figure 15, from where one can see that $A$ is proportional to $ln(B)$ up to a certain field level, and with the $B$ field in 5~30 mT range, parameter $A$ changes in the 20~10 μΩ range, consistent with the experimental fitting shown in [20]. The change of $U$ between 5 and 30 mT is quite small, ~0.02meV, also consistent with the results in [20].

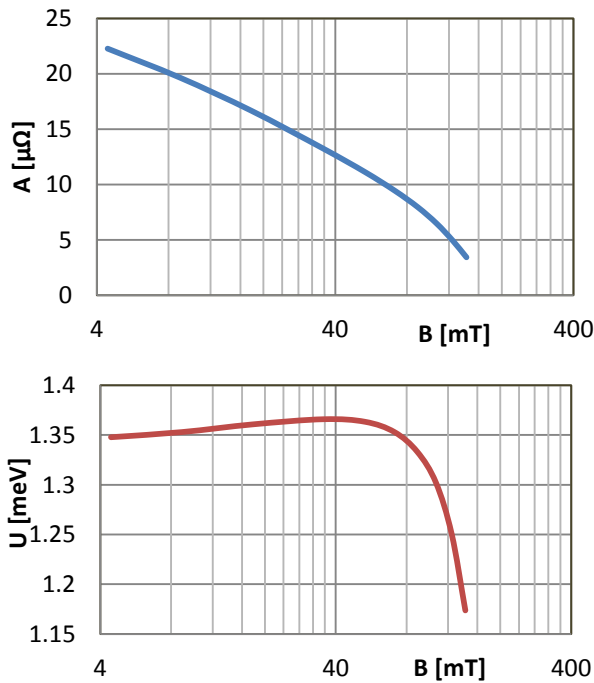

Figure 15. Dependencies of $A$ and $U$ on rf field $B$, fitted to $R_s(T)=Aexp(-U/kT)$ using data in Figure 9.

The correspondence of the recent "high Q" data to the predictions of Xiao's extension of M-B theory of SRF surface impedance is striking. Significant further study is needed to examine experimentally the temperature dependence of the loss mechanisms present to further test the theoretical predictions. This has begun and will be reported elsewhere.

## DISCUSSION

Since the extended theory is a treatment of an "ideal" BCS superconductor, one may interpret the observed increasing $Q$ as the way "good" niobium should be expected to perform. An implication is that the "normal" niobium to which the community is presently accustomed is actually "polluted" in some way, at least within the RF penetration depth, in a way which contributes very common additional losses [9].

SRF losses have been studied using temperature mapping systems [13, 21], with thermal feedback model [22] and localized quench spot [13, 21], using topographic profile with surface roughness model [23] and field enhancement model [24], and/or considering the oxygen diffusion from X-ray photoelectron spectroscopy with oxygen pollution model [25]. Other possible losses, including the normal conducting core [26] and the vortex [27], could also address mechanism for rf losses additional to the inherent BCS losses considered in this extension theory and give possible explanations to the $Q$ slope of "normal" niobium cavities. Local "normal precipitates" that are superconducting by proximity effect until the local field exceeds a specific value also may give an explanation to the "normal" $Q$ slope [28]. As the local surface magnetic field increases, more of the penetration depth's volume exceeds this value, so the $R_s$ contributed by these localized, but normal, precipitates would effectively increases with field amplitude, contributing a middle field $Q$ decrease.

Clarification of such mechanisms and the engineering of processes to avoid them would seem to be quite worthy undertakings. Success at this would enable very significant improvements in the economy of SRF-based accelerator construction and operation. The cost optimization of cryoplant capital and operating expenses together with accelerator systems might change considerably if these theoretical predictions and recent low loss data can be generalized. For example, with recently demonstrated L-band elliptical $Q$ values increasing from $1\times10^{10}$ to $5\times10^{10}$ under reasonably high fields, CW accelerator applications with loss much higher than the static loss at 2 K operating temperature, the heat load of the cryoplant might see reductions approaching 75%.

Further down the road, since the theory is general to all BCS superconductors, one might anticipate even further cryogenic cost benefits from the use of higher-$T_c$ materials.

## SUMMARY

A field-dependent derivation of M-B theory of the RF surface impedance of BCS superconductors has been introduced with no need of any additional parameters. Despite the complexity of the mathematical expressions, numerical calculation results show a good correspondence to recent high-$Q$ experimental results. The attractive $Q$-increase with peak RF fields up to 80 mT is explained based on the quasi-particle scattering procedure that may have a decreasingchance to occur from a lower energy state to a higher energy state with energy difference larger than the photon energy, and the averaged reduced opportunities for transitions comes from the angle-dependent modified single particle distribution function. A parametric sensitivity survey has been performed to obtain the field-dependent RF surface impedance sensitivity to BCS material parameters in hopes of supporting increased insight into performance optimization strategies.

# ACKNOWLEDGEMENT

Authored by Jefferson Science Associates, LLC under U.S. DOE Contract No. DE-AC05-06OR23177, and by Brookhaven Science Associates, LLC under Contract No. DE-AC02-98CH10886. The U.S. Government retains a non-exclusive, paid-up, irrevocable, world-wide license to publish or reproduce this manuscript for U.S. Government purposes. One of the authors would like to thank Drs. Ilan Ben-Zvi and Sergey Belomestnykh for their comments during the preparation of this manuscript.

\# binping@bnl.gov